\documentclass[useAMS,usenatbib]{mn2e}
\usepackage{epsfig}
\usepackage{amsmath}

\newcommand{\be}{\begin{equation}}
\newcommand{\beq}{\begin{equation}}
\newcommand{\ba}{\begin{eqnarray}}
\newcommand{\ee}{\end{equation}}
\newcommand{\eeq}{\end{equation}}
\newcommand{\ea}{\end{eqnarray}}

\def\lsim{~\rlap{$<$}{\lower 1.0ex\hbox{$\sim$}}}

\def\gsim{~\rlap{$>$}{\lower 1.0ex\hbox{$\sim$}}}

\title[Quasar HII Regions]{The Impact of a Percolating IGM on Redshifted 21\,cm Observations of Quasar HII Regions.}

\author[Geil \& Wyithe]{Paul M. Geil \& J. Stuart B. Wyithe\\School of Physics, University of Melbourne, Parkville, Victoria, Australia\\Email: pgeil@physics.unimelb.edu.au, swyithe@physics.unimelb.edu.au}

\begin{document}


\maketitle

\label{firstpage}
\begin{abstract}
We assess the impact of inhomogeneous reionization on detection of HII regions surrounding luminous high redshift quasars using planned low frequency radio telescopes. Our approach is to implement a semi-numerical scheme to calculate the 3-dimensional structure of ionized regions surrounding a massive halo at high redshift, including the ionizing influence of a luminous quasar. As part of our analysis we briefly contrast our scheme with published semi-numerical models. We calculate mock 21\,cm spectra along the line of sight towards high redshift quasars, and estimate the ability of the planned Murchison Widefield Array to detect the presence of HII regions. The signal-to-noise for detection will drop as the characteristic bubble size grows during reionization because the quasar's influence becomes less prominent. However, quasars will imprint a detectable signature on observed 21\,cm spectra that is distinct from a region of typical IGM. At epochs where the mean hydrogen neutral fraction is $\approx30\%$ or greater we find that neutral gas in the IGM surrounding a single quasar will be detectable (at a significance of $5\sigma$) within 100 hour integrations in more than 50\% of cases. 1000 hour integrations will be required to detect a smaller neutral fraction of 15\% in more than 50\% of cases. A highly significant detection will be possible in only 100 hours for a stack of 10 smaller 3 proper Mpc HII regions. The accurate measurement of the global average neutral fraction ($\langle x_{\textrm{HI}}\rangle$) will be limited by systematic fluctuations between lines of sight for single HII regions. We estimate the accuracy with which the global neutral fraction could be measured from a single HII region to be 50\%, 30\% and 20\% for $\langle x_{\textrm{HI}}\rangle \approx$ 0.15, 0.3 and 0.5 respectively.
\end{abstract}

\begin{keywords}
cosmology: diffuse radiation, theory -- galaxies: high redshift, intergalactic medium
\end{keywords}
 
\section{Introduction}
\label{intro}
The reionization of cosmic hydrogen is commonly believed to have been due to UV photons produced by the first stars and quasars and was an important milestone in the history of the Universe (e.g. \citeauthor{bl2001} \citeyear{bl2001}). The recent discovery of very distant quasars has allowed detailed absorption studies of the state of the high redshift intergalactic medium (IGM) at a time when the Universe was less than a billion years old (e.g. \citeauthor{fan2006} \citeyear{fan2006}; \citeauthor{white2003} \citeyear{white2003}).  Several studies have used the evolution of the ionizing background inferred from these spectra to argue that the reionization of cosmic hydrogen was completed just beyond $z = 6$ \citep{fan2006, gnedin2006, white2003}. However, other authors have claimed that the evidence for this rapid change is the result of an incorrect choice for the distribution of overdensities in the IGM \citep{becker2007}.

Several of the most distant quasars exhibit complete Gunn-Peterson troughs, the red edges of which do not extend as far as the redshifted Ly$\alpha$ wavelength.  The simplest interpretation of this feature is the presence of an HII region in a partially neutral IGM (e.g. \citeauthor{cen2000} \citeyear{cen2000}). This idea has been used to argue that there is a significant fraction of neutral hydrogen in the IGM beyond $z\approx6$ \citep{wl2004a,mesinger2004, wlc2005}, with a lower limit on the hydrogen neutral fraction that is as many as two orders of magnitude larger than constraints provided by direct absorption studies. On the other hand, more detailed numerical modeling has shown that this interpretation is uncertain (\citeauthor{lidz2007a} 2007a; \citeauthor{bolton2007} \citeyear{bolton2007}), and that the observed spectra could either be due to an HII region or a classical proximity zone.

The difficulties in interpreting spectra of high redshift quasars arise because the large optical depth for Ly$\alpha$ absorption in a neutral IGM limits its effectiveness in probing reionization at redshifts beyond $z\approx6$.  A better probe of the reionization history involves the redshifted 21\,cm emission from neutral hydrogen in the IGM. Several probes of the reionization epoch in redshifted 21\,cm emission have been suggested. These include observation of the emission as a function of redshift averaged over a large area of sky. This observation would provide a direct probe of the evolution in the neutral fraction of the IGM, and is referred to as the global step \citep{shaver1999,gnedin2004}. A more powerful probe will be provided by observation of the power spectrum of fluctuations together with its evolution with redshift.  This observation would trace the evolution of neutral gas with redshift as well as the topology of the reionization process (e.g. \citeauthor{tozzi2000} \citeyear{tozzi2000}; \citeauthor{furl2004b} \citeyear{furl2004b}; \citeauthor{loeb2004} \citeyear{loeb2004}; \citeauthor{wm2007} \citeyear{wm2007}; \citeauthor{iliev2006} \citeyear{iliev2006}). Finally, observation of individual HII regions will probe quasar physics as well as the evolution of the neutral gas \citep{wl2004b,kohler2005}. \cite{kohler2005} have generated synthetic spectra using cosmological simulations and conclude that quasar HII regions will provide the most prominent individual cosmological signals. Recent work by \cite{datta2007b} has focused on the detection of ionized bubbles in redshifted 21\,cm maps using a visibility based formalism. Their results suggest it may be possible to blindly detect spherical HII regions of radius$\gsim$ 22 comoving Mpc during the epoch of reionization.

Various experiments are planned to measure 21\,cm emission from the pre-reionization IGM, including the Low Frequency Array\footnote{http://www.lofar.org/} (LOFAR; \citeauthor{valdes2006} \citeyear{valdes2006}) and the Murchison Widefield Array\footnote{http://www.haystack.mit.edu/ast/arrays/mwa/} (MWA). The MWA is being constructed on a very radio quiet site in Western Australia and will consist of 512 tiles, each containing 16 cross dipoles, which operate over the range 80-300\,MHz. The tiles will be spread to cover baselines of up to 1.5\,km. In collecting area the MWA would be comparable to the Very Large Array (VLA). However, the correlator/receiver system will be able to handle a much larger number of channels than are available on the low frequency VLA. In this paper we focus on the observation of individual HII regions around known quasars with emphasis on the capabilities of the MWA. In particular we address the influence of the percolating IGM on the redshifted 21\,cm signal from the neutral hydrogen surrounding a quasar HII region. Our analysis extends ideas presented by \cite{wlb2005} by including calculations of a realistic structure for ionization of the IGM, as well as estimates of the achievable signal-to-noise.

We discuss a simple detection strategy consisting of a spectral fit to the 21\,cm signal along the line of sight towards a known high redshift quasar. The following \textit{back-of-the-envelope} calculation gives a comparitive gauge of the statistical significance required by this strategy relative to a blind search for individual HII regions, embedded in a partially ionized IGM.

For a blind search, the number of independent samples within a single MWA field is given by
\begin{equation}
N_{\rm s} = \frac{V_{\rm MWA}}{V_{\rm q}(z,R_{\rm q},R_{\rm b})} \sim \frac{V_{\rm MWA}}{\pi R^2_{\rm b} \times 2R_{\rm q}},
\end{equation}
where $V_{\rm MWA}$ is the comoving volume of the MWA field and $V_{\rm q}(z,R_{\rm q},R_{\rm b})$ is the intersection of the synthesized beam (of comoving radius $R_{\rm b}$) with the HII region (of comoving radius $R_{\rm q}$). For a 4.5 proper Mpc region at $z=6.5$ using a synthesized beam of width $\theta_{\rm b} = 3.2^{\prime}$ we obtain $V_{\rm q} \sim 10^4$\,comoving Mpc$^3$. Together with $V_{\rm MWA} \sim \pi 4700^2 \times 260 \sim 10^{10}$\,comoving Mpc$^3$ \citep{bowman2006}, we have $N_{\rm s} \sim 10^{6}$.

Given a signal-to-noise level of $n\sigma$ for detection of the mean IGM, the number of \textit{false positive} detections per MWA field is 
\begin{equation}
N_{\rm false} \sim N_{\rm s}\,\rm{erfc}\left(\frac{n}{\sqrt{2}\sigma}\right),
\end{equation}
where $\textrm{erfc}(x)$ is the complimentary error function. For example, if the mean IGM can be detected with a signal-to-noise level of $2\sigma$, we expect $\sim10^4$ false detections per field, but only one true detection of a quasar HII region. The chance of obtaining only one false detection requires $n \approx 5$. Therefore, in order to reduce the chance of a false detection and have confidence in the reality of the HII region we require $N_{\rm false}\ll1$ or $n\gg5$. This level of significance will probably not be possible with the expected sensitivity of the MWA which will make a blind search impractical.

On the other hand, by making an observation around a quasar, we are able to take a Bayesian approach by utilizing the likelihood of detection given that an HII region is known \textit{a priori} to be present. In this case a detection at a given significance level is not diluted by the possibility of false detections. We stress that even when the impact of the quasar is no longer dominant, the method of detecting the 21\,cm signal \textit{a posteriori} along the line of sight to a known quasar will still succeed, because neutral gas is known \textit{a priori} to be absent in a large volume surrounding the quasar.

In addition to its physical configuration, the sensitivity of a radio telescope is limited due to difficulties in achieving a perfect calibration and to problems introduced by foregrounds and the ionosphere. There has been significant discussion in the literature regarding foregrounds. While fluctuations due to foregrounds are orders of magnitude larger than the reionization signal (e.g. \citeauthor{dimatteo2002} \citeyear{dimatteo2002}; \citeauthor{oh2003} \citeyear{oh2003}), foreground spectra are anticipated to be smooth. Since the reionization signature includes fluctuations in frequency as well as angle, they can be removed through continuum subtraction (e.g. \citeauthor{gnedin2004} \citeyear{gnedin2004}; \citeauthor{wang2006} \citeyear{wang2006}), or using the differences in symmetry from power spectra analysis \citep{morales2004,zaldarriaga2004}. In this paper we concentrate on theoretical limitations of the MWA and the impact of the inhomogeneous reionization of the IGM. The practical difficulties introduced by calibration and foreground removal etc. need to be addressed by full simulations of interferometric observations, and ultimately with early data sets from new facilities.

We begin by describing the inclusion of stellar reionization and a luminous quasar within a semi-numerical scheme to produce realistic fluctuations in the ionized gas distribution (\S\,\ref{model}). We then use this model to describe the evolution of the ionization structure of the IGM surrounding a massive quasar host (both with and without the presence of a quasar HII region) in \S\,\ref{evolution}, before discussing the detectability of these regions in \S\,\ref{detect}. We present our conclusions in \S\,\ref{conclusion}.  Throughout this paper we adopt the set of cosmological parameters determined by WMAP \citep{spergel2007} for a flat $\Lambda$CDM universe.

\section{Semi-Numerical Model for the Growth of Quasar HII Regions in a Percolating IGM}
\label{model}
In this section we describe our model for the reionization of a 3-dimensional volume of the IGM using a semi-numerical scheme that is similar in spirit to the models described by \cite{zahn2007} and \cite{mesinger2007}, based upon \cite{furl2004a}. We begin with a brief description of the generation of the Gaussian initial conditions in the density field surrounding a massive quasar host halo at $z\sim 7$. We then describe a semi-analytic model for the density dependent reionization of the IGM, including the influence of a nearby quasar, before combining the numerical realization of the density field with our semi-analytic calculations to generate 3-dimensional realizations of the ionization state of the IGM.

\subsection{Construction of density field surrounding a massive halo}
We simulate the linear density field $\delta(\mathbf{k})$ inside a periodic, comoving, cubic region of volume $V=L^{3}$, by calculating the density contrast in Fourier space $\hat{\delta}(\mathbf{k})$ corresponding to a $\Lambda$CDM power spectrum at a specified redshift. The power spectrum used has been normalized using $\sigma_{8}$ and is given by $P(k,z) \propto k^{n}T^{2}(k)D^{2}(z)$ where $T(k)$ is an analytic approximation of the present day transfer function \citep{bardeen1986}, $n=0.96$ is the primordial power spectral index \citep{spergel2007} and $D(z)$ is the linear growth factor between redshift $z$ and the present.

Numerically, this procedure is carried out by laying down a cubic grid of size $N^3$ and approximating the continuous Fourier relationship between $\delta(\mathbf{x})$ and $\hat{\delta}(\mathbf{k})$ with its discrete form,
\begin{equation}
\hat{\delta}(\mathbf{k}) = \frac{V}{N^{3}}\sum_{\mathbf{x}} \delta(\mathbf{x})e^{-i\mathbf{k}\cdot\mathbf{x}}.
\end{equation}
Each Fourier density mode is calculated by sampling the power spectrum at each of the finite set of wavenumbers $\mathbf{k}=(l,m,n)k_{0}$, where $l,m,n\in\{-N/2+1,-N/2+2,\dots,N/2\}$ and $k_{0}=2\pi/L$. In order to realize the Gaussianity of $\delta(\mathbf{x})$ we draw both the real and complex components of $\hat{\delta}(\mathbf{k})$ from a zero-mean normal distribution with variance $\sigma^{2}=P(k,z)V/2$. Furthermore, we enforce Hermiticity upon $\hat{\delta}(\mathbf{k})$ to ensure $\delta(\mathbf{x})$ is real. All 7 vertices corresponding to the positive frequency Nyquist modes must be real, and in order to produce a zero-mean density field the DC mode $\hat{\delta}(\mathbf{k}=0)$ is set to zero. The 3-dimensional inverse Fourier transform of $\hat{\delta}(\mathbf{k})$ gives an Eulerian realization of $\delta(\mathbf{x})$ filtered on a scale $\Delta=L/N$ (e.g. \citeauthor{sirko2005} \citeyear{sirko2005}).

To study the impact of a high redshift quasar on the ionization state of its surrounding IGM we identify the approximate location of a collapsed halo of mass $M$ by smoothing the linear density field using a real-space spherical top-hat filter of radius $R_{\rm f}=(3M/4\pi\bar{\rho_{0}})^{1/3}$, where $\bar{\rho_{0}}=3\Omega_{\rm m}H_{0}^{2}/8\pi G$ is the present mean matter density. This is equivalent to multiplying the Fourier coefficients of $\hat{\delta}(k)$ by the normalized kernel $\hat{W}(k) = 3j_{1}(kR_{\rm f})/kR_{\rm f}$, where $j_{1}(x)$ is the first-order spherical Bessel function. Density peaks are considered to have collapsed if $\delta(\mathbf{x},z)> \delta_{\rm c}$, where $\delta_{\rm c}\approx1.686$ is the critical linear overdensity of a spherical top-hat density perturbation. Note that the field is linearly evolved to $z$ before the $\delta_{\rm c}$ filter is applied. We fiducially set $M=10^{12}$\,$M_{\odot}$ as the minimum value for the masses of quasar host halos at $z\approx6.5$ with proximity zones of $R\approx30$ comoving Mpc \citep{fan2006}.

\subsection{Density dependent model of global reionization}
Our semi-analytic model to compute the relation between the local dark matter overdensity and the reionization of the IGM is based on the model described by \cite{wl2007} and \cite{wm2007}. Here we summarize the main features of the model and refer the reader to those papers for more details.

Any model for the reionization of the IGM must describe the relation between the emission of ionizing photons by stars in galaxies and the ionization state of the intergalactic gas. This relation is non-trivial as it depends on various internal parameters (which may vary with galaxy mass), such as the fraction of gas within galaxies that is converted into stars and accreting black holes, the spectrum of the ionizing radiation, and the escape fraction of ionizing photons from the surrounding interstellar medium as well as the galactic halo and its immediate infall region [see \cite{loeb2006} for a review]. The relation also depends on intergalactic physics.  In overdense regions of the IGM, galaxies will be overabundant because small-scale fluctuations need to be of lower amplitude to form a galaxy when embedded in a larger scale overdensity \citep{mo1996}. On the other hand, the increase in the recombination rate in overdense regions counteracts this {\it galaxy bias}. The process of reionization also contains several layers of feedback.  Radiative feedback heats the IGM and results in the suppression of low-mass galaxy formation \citep{efstathiou1992,thoul1996,quinn1996,dijkstra2004}. This delays the completion of reionization by lowering the local star formation rate, but the effect is counteracted in overdense regions by the biased formation of massive galaxies. 

The evolution of the ionization fraction by mass $Q_{\delta,R}$ of a particular region of scale $R$ with overdensity $\delta$ (at observed redshift $z_{\rm obs}$) may be written as
\begin{eqnarray}
\label{history}
\nonumber
\frac{dQ_{\delta,R}}{dt} &=& \frac{N_{\rm ion}}{0.76}\left[Q_{\delta,R} \frac{dF_{\rm col}(\delta,R,z,M_{\rm ion})}{dt} \right.\\
\nonumber
&&\hspace{5mm}+ \left.\left(1-Q_{\delta,R}\right)\frac{dF_{\rm col}(\delta,R,z,M_{\rm min})}{dt}\right]\\
&-&\alpha_{\rm B}Cn_{\rm H}^0\left[1+\delta\frac{D(z)}{D(z_{\rm obs})}\right] \left(1+z\right)^3Q_{\delta,R},
\end{eqnarray}
where $N_{\rm ion}$ is the number of photons entering the IGM per baryon in galaxies, $\alpha_{\rm B}$ is the case-B recombination coefficient, $C$ is the clumping factor (which we assume, for simplicity, to be a constant value of 2) and $D(z)$ is the growth factor between redshift $z$ and the present. The production rate of ionizing photons in neutral regions is assumed to be proportional to the collapsed fraction $F_{\rm col}$ of mass in halos above the minimum threshold mass for star formation ($M_{\rm min}$), while in ionized regions the minimum halo mass is limited by the Jeans mass in an ionized IGM ($M_{\rm ion}$). We assume $M_{\rm min}$ to correspond to a virial temperature of $10^4$\,K, representing the hydrogen cooling threshold, and $M_{\rm ion}$ to correspond to a virial temperature of $10^5$\,K, representing the mass below which infall is suppressed from an ionized IGM \citep{dijkstra2004}. In a region of comoving radius $R$ and mean overdensity $\delta(z)=\delta D(z)/D(z_{\rm obs})$ (specified at redshift $z$ instead of the usual $z=0$), we use the extended Press-Schechter \citep{ps1974} model \citep{bond1991} to calculate the relevant collapsed fraction
\begin{equation}
F_{\rm col}(\delta,R,z) = \mbox{erfc}{\left(\frac{\delta_{\rm c}-\delta(z)}{\sqrt{2[\sigma^2_{\rm gal}-\sigma^2(R)]}}\right)},
\end{equation}
where $\mbox{erfc}(x)$ is the complimentary error function, $\sigma^2(R)$ is the variance of the density field smoothed on a scale $R$, and $\sigma^2_{\rm gal}$ is the variance of the density field smoothed on a scale $R_{\rm gal}$, corresponding to a mass scale of $M_{\rm min}$ or $M_{\rm ion}$ (both evaluated at redshift $z$ rather than at $z=0$).

Equation~(\ref{history}) may be integrated as a function of $\delta$.  At a specified redshift this yields the filling fraction of ionized regions within the IGM on various scales $R$ as a function of overdensity. We find that this model predicts the sum of astrophysical effects to be dominated by galaxy bias, and that as a result overdense regions are reionized first. This leads to the growth of HII regions via a phase of percolation during which individual HII regions overlap around clustered sources in overdense regions of the universe. We may also calculate the corresponding 21\,cm brightness temperature contrast 
\begin{equation} T(\delta,R) =
22\mbox{ mK }(1-Q_{\delta,R})\left(\frac{1+z}{7.5}\right)^{1/2}\left(1+\delta\right).
\end{equation} 
In this paper we ignore the enhancement of the brightness temperature fluctuations due to peculiar velocities in overdense regions \citep{bharadwaj2005,bl2005}. Peculiar velocities were included in the semi-numerical model of \cite{mesinger2007} who found their effect to be small on scales $\sim 10$ comoving Mpc.

\subsection{The ionization field}
Having determined the value of the ionized fraction $Q_{\delta,R}$ as a function of overdensity $\delta$ and smoothing scale $R$, we may now construct the ionization field.  We employ a similar filtering algorithm to that of \cite{mesinger2007} to determine the ionization state at each grid position. This is done by repeatedly filtering the linear density field on scales in the range from $L$ to $L/N$ at logarithmic intervals of width $\Delta R_{\textrm{f}}/R_{\textrm{f}}=0.1$. For all filter scales, the ionization state of each grid position is determined using $Q_{\delta,R}$ and deemed to be fully ionized if $Q_{\delta,R}\geqslant 1$. All voxels within a sphere of radius $R$ centered on these positions are flagged and assigned $Q_{\delta,R}=1$, while the remaining non-ionized voxels are assigned an ionized fraction of $Q_{\delta,R_{\textrm{f,min}}}$, where $R_{\textrm{f,min}}=L/N$ corresponds to the smallest smoothing scale. A voxel forms part of an HII region if $Q_{\delta,R}>1$ on any scale $R$.

\subsection{Inclusion of quasars in the semi-numerical scheme}
\label{quasar}
In a region of radius $R$ containing a quasar, the cumulative number of ionizations per baryon will be larger than predicted from equation~(\ref{history}). We include quasars in our scheme by first computing the fraction of the IGM within a region of radius $R$ centered on the position $\mathbf{x}$ that has already been reionized by stars [using equation (\ref{model})], we then add an additional ionization fraction equal to the quasar's contribution 
\begin{equation}
Q_{\rm q}(\mathbf{x}) = \left(\frac{|\mathbf{x}-\mathbf{x}_{\rm q}|}{R_{\rm q}}\right)^{-3},
\end{equation}
where $R_{\rm q}$ is the radius of the HII region centered on $\mathbf{x}_{\rm q}$ that would have been generated by the quasar alone in a fully neutral IGM. In contrast to stellar ionization the quasar contribution comes from a point source, and so the contribution to $Q$ originates from a single voxel only (rather than all voxels within $|\mathbf{x}-\mathbf{x}_{\rm q}|$). Following this addition we filter the ionization field as described in \S\,2.3.

\section{The Evolution of HII Regions in a Percolating IGM}
\label{evolution}
In this section we describe one example of the evolution of the IGM during the reionization epoch. We first show a model for the evolution of ionized regions due to stellar ionization alone in \S\,3.1, before adding the presence of a luminous quasar in \S\,3.3.

\begin{figure*}
\includegraphics[width=17cm]{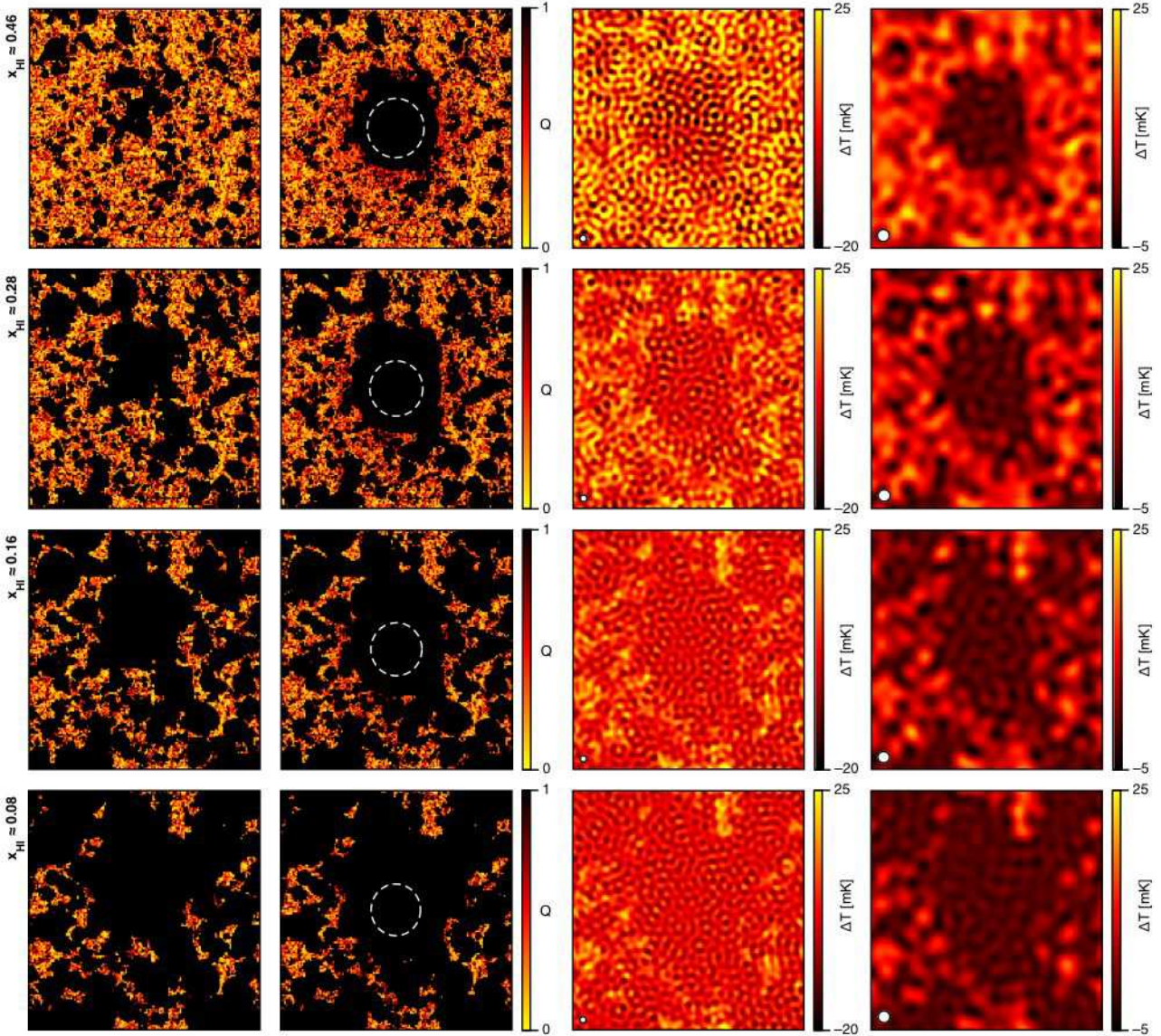} 
\caption{ Ionization maps of the IGM showing the evolution of global ionization fraction, $Q$, with and without the ionizing contribution of a quasar. The slices shown correspond to epochs when the global, volume-weighted neutral fraction is $\langle x_{\textrm{HI}}\rangle \approx 0.46,$ 0.28, 0.16 and 0.08 (\textit{top} to \textit{bottom}) corresponding to $z = 7.5,$ 7.0, 6.7 and 6.5 respectively. Each slice has a comoving side length of 219\,$h^{-1}$\,Mpc and is 0.86\,$h^{-1}$ comoving Mpc deep. The ionization maps were computed within boxes of dimension 256$^{3}$. {\em Left:} Evolution with no quasar. {\em Middle left:} Corresponding ionization maps with an embedded $R_{\textrm{q}} \sim 34$ comoving Mpc quasar HII region. Dashed circles represent the HII region that would have been generated by the quasar alone in a \textit{fully neutral} IGM. {\em Middle right:} Corresponding synthetic maps with $\theta_{\rm b} \approx 3.2\,^{\prime}$ (corresponding to central peak to first null $\Delta\theta \approx 5.5\,^{\prime}$) including quasar HII region using the MWA and 1000\,hr integrations. {\em Right:} Corresponding synthetic maps with $\theta_{\rm b} \approx 5.8\,^{\prime}$ (corresponding to central peak to first null $\Delta\theta \approx 10.0\,^{\prime}$) including quasar HII region using the MWA and 1000\,hr integrations.}
\label{fig1}
\end{figure*}

\subsection{Fiducial model for reionization}
\begin{figure}
\includegraphics[width= 8.3cm]{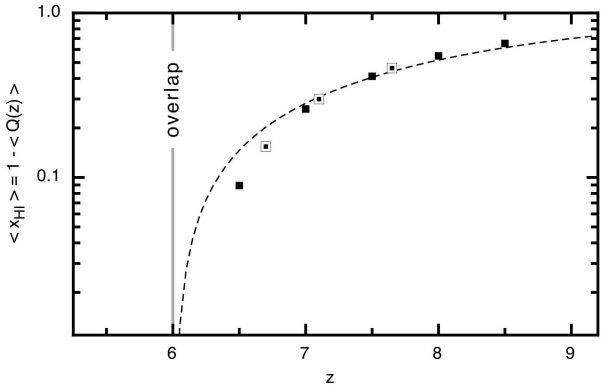} 
\caption{Global, mass-weighted neutral fraction as a function of redshift. The dashed line shows the evolution of neutral fraction as expected from the density dependent model described in \S\,2.2 (for an overlap at $z \approx 6$). The data show the mass-weighted globally averaged neutral fraction from our simulations. Hollow squares indicate the approximate epoch for which simulations were used to obtain the mock line of sight spectra shown in Figure \ref{fig4}.}
\label{fig2}
\end{figure}

Throughout this paper we consider a model in which the mean IGM is reionized at $z=6$ \citep{fan2006,gnedin2006,white2003}. In this model we assume that star formation proceeds in halos above the hydrogen cooling threshold in neutral regions of IGM. In ionized regions of the IGM star formation is assumed to be suppressed by radiative feedback (see \S\,\ref{model}). We compute the 3-dimensional structure for this model using the prescription outlined in \S\,\ref{model}. In the left-hand panel of Figure~\ref{fig1} we present snapshots of the ionization fraction  $Q(\textbf{x})$ for slices through a realization of this model when the mean, volume-weighted neutral fraction of the IGM is $\approx 0.5,$ 0.3, 0.15 and 0.1 corresponding to redshifts $z=7.5,$ 7.0, 6.7 and 6.5 respectively. The simulation presented corresponds to a  linear density field of resolution $256^{3}$, with a comoving side length of 219\,$h^{-1}$\,Mpc. Each slice is 0.86\,$h^{-1}$ comoving Mpc deep. Since the box is centered on a massive halo the simulation shows a large HII region growing in the center of the field. As reionization progresses we see this central region merge with adjacent HII regions. This process of overlap is seen in detailed numerical simulations (e.g. \citeauthor{mcquinn2007} \citeyear{mcquinn2007}) and leads to an irregular structure of neutral gas filaments late in the reionization era. The overlap of some HII regions, combined with the fact that our model does not allow for the ionizing flux from one source to contribute to the ionization due to another results in a violation of photon conservation (e.g. \citeauthor{zahn2007} \citeyear{zahn2007}). Although this effect is expected to be small during the early stages of reionization when there are fewer HII region mergers, it does lead to more severe discrepancies at later times when the global ionization fraction approaches unity. Figure~\ref{fig2} illustrates this effect by plotting the global neutral fraction as a function of redshift from our simulations. The dashed line shows the evolution of mean neutral fraction by mass as expected from the analytic model which provides the input density dependent ionization fraction. We find that our semi-numerical scheme yields results that conserve photons at the $\sim15\%$ level for neutral fractions $\langle x_{\textrm{HI}}\rangle \gsim\;0.2$.

\subsection{Comparison with other semi-numeric models}
We note that our scheme correctly assigns the ionization fraction on scales that are not resolved by the simulation cube. This is not necessarily the case in similar models that approximate the ionization field as a two-phase medium (e.g. \citeauthor{mesinger2007} \citeyear{mesinger2007}), although a two-phase approximation may be justified depending upon the resolution of the simulation. Implementing our scheme with increased resolution would improve the accuracy and dynamical range of our simulations. However, we maintain a high resolution \textit{relative} to the synthesized beam. The assignment of regions as either fully neutral or fully ionized may also lead to an overestimate of the brightness temperature contrast across ionization fronts which are observed at finite resolution. Our implementation achieves approximately the correct relation between neutral fraction and epoch over most of the reionization history and produces soft transitions at the edges of large HII regions by smoothing over HII regions not resolved by the simulations. However, as noted by \cite{mesinger2007} a two-phase approximation is appropriate for the purposes of describing the morphology and evolution of HII regions. We also point out that there are several features not incorporated in our model. These include the galaxy distribution (which will enhance fluctuations on small scales due to Poisson noise) and the velocity field of the gas (see \citeauthor{mesinger2007} \citeyear{mesinger2007}). However these additional considerations are not important on scales larger than $10$ comoving Mpc, and so do not influence our results regarding HII regions surrounding high redshift quasars. 

\subsection{The influence of a quasar in the fiducial model for reionization}
We next add quasar luminosity to the simulations shown in the left-hand panels of Figure \ref{fig1} using the prescription outlined in \S\,\ref{quasar}. The value of $R_{\rm q}$ is subject to large uncertainties, including the quasar duty cycle, lifetime and luminosity. Assuming a line of sight ionizing photon emission rate of $\dot{N}_{\gamma}/4\pi$ photons per second per steradian, a uniform IGM and neglecting recombinations, the line of sight extent of a quasar's HII region (from the quasar to the \textit{front} of the region) observed at time \textit{t} is \citep{white2003}
\begin{equation}\label{RqWhite}
R_{\rm q} \approx 4\,\textrm{pMpc}\;x_{\rm HI}^{-1/3}\left(\frac{\dot{N}_{\gamma}}{10^{57}}\frac{t_{\rm age}}{10^7 \textrm{yr}}\right)^{1/3}\left(\frac{1+z}{7.5}\right)^{-1},
\end{equation}
where $x_{\rm HI}$ is the neutral fraction and $t_{\rm age}=t-R(t_{\rm age})/c$ is the age of the quasar corresponding to the time when photons that reach $R_{\rm q}$ at time $t$ were emitted.

We show examples of $R_{\rm q} \approx34$ comoving Mpc, consistent with the lower limit around the luminous SDSS quasars at $z>6$ \citep{fan2006}. The second column of Figure~\ref{fig1} shows the same realization of the IGM presented in the first column, but now includes the influence of a luminous quasar hosted by the central massive halo.  As before, each slice has a comoving side length of 219\,$h^{-1}$\,Mpc and is 0.86\,$h^{-1}$ comoving Mpc deep. Note that while the evolution of the ionization due to stars has been followed self-consistently in these simulations, we have assumed the quasar to contribute the {\em same} proper volume of ionization at each redshift. The apparent evolution in the size of the HII region surrounding this quasar is therefore due to the additional contributions from nearby galaxies to the HII region (and not due to the evolution of IGM density since the slices are shown in comoving units). 

The quasar HII region appears spherical and distinct early in the reionization era when the global neutral fraction is still large, and the typical size of galaxy driven HII regions is small (i.e. $z\ga8$ in this model). However, the shape of the HII region becomes highly distorted once the typical galaxy (or galaxy cluster) generated HII region becomes large late in the reionization epoch \citep{furl2004b}. Indeed, once the size of the typical HII region exceeds that generated by the quasar, the quasar HII region ceases to be an identifiable feature. We note that a bandwidth of $\Delta\nu$, centered on $z$, corresponds to the redshift interval $\Delta z = \Delta\nu(1+z)^2/1420$\,MHz, implying that a 20\,MHz band (the approximate frequency range of our synthetic spectra) has a redshift extent of $\Delta z \approx 0.8$. We expect significant evolution of the 21\,cm signal over this redshift interval very close to overlap. However, given the other uncertainties in our modeling we ignore such evolution over this bandwidth in our simulations.

\section{Detectability of HII regions}
\label{detect}
In the remainder of this paper we discuss estimates for the detectability of the predicted signature of quasar HII regions in 21\,cm emission. In particular, we discuss the sensitivity of the MWA to the HII regions surrounding known high redshift quasars in a percolating IGM.

\subsection{Sensitivity to the 21\,cm signal}
\label{sensitivity}
Radio interferometers measure a frequency-dependent, complex visibility $V(\mathbf{U},\nu)$ (Jy) for each frequency channel and baseline $\mathbf{U}$ in their configuration. The measured visibility is, in general, a linear combination of signal, foreground and system noise,
\begin{equation}
\label{vis}
V(\mathbf{U},\nu) = V_{\rm S}(\mathbf{U},\nu) + V_{\rm F}(\mathbf{U},\nu) + V_{\rm N}(\mathbf{U},\nu),
\end{equation}
where $V_{\rm S}$ is the signal, $V_{\rm F}$ the contribution due to foreground sources and $V_{\rm N}$ the system noise. The signal of interest in this work is redshifted 21\,cm emission. In order to discuss the response of a phased array to the brightness temperature contrast of the 21\,cm emission from the IGM we must i) quantify the instrumental noise-induced uncertainty in each visibility $\Delta V_{\rm N}$ and ii) address foreground subtraction. The rms noise fluctuation per visibility per frequency channel is given by
\begin{equation}
\label{radiometer}
\Delta V_{\rm N} = \frac{2k_{\rm B}T_{\rm sys}}{A_{\rm eff}\sqrt{t_{\mathbf{U}}\Delta\nu}},
\end{equation}
where $T_{\rm sys}$ is the system temperature (K), $A_{\rm eff}$ the effective area of one antenna (m$^{2}$), $t_{\mathbf{U}}$ is the integration time for that visibility (s), $\Delta\nu$ is the frequency bin width (Hz) and $k_{\rm B}$ is the Boltzmann constant. Instrumental noise is uncorrelated in the frequency domain. As shown by \cite{mcquinn2006}, the average integration time $t_{\mathbf{U}}$ that an array observes the visibility $\mathbf{U}$ is
\begin{equation}
\label{tU}
t_{\mathbf{U}} \approx \frac{A_{\rm eff}t_{\rm int}}{\lambda^{2}}n(\mathbf{U}),
\end{equation}
where $t_{\rm int}$ is the total integration time and $n(\mathbf{U})$ is the number density of baselines that can observe the visibility $\mathbf{U}$. The transverse wavenumber $k_{\perp}$ is given by $k_{\perp} = 2\pi |\mathbf{U}|/r_{\rm em}(z)$, where $r_{\rm em}(z)$ is the proper distance to the point of emission.

\begin{figure}
\includegraphics[width= 8.5cm]{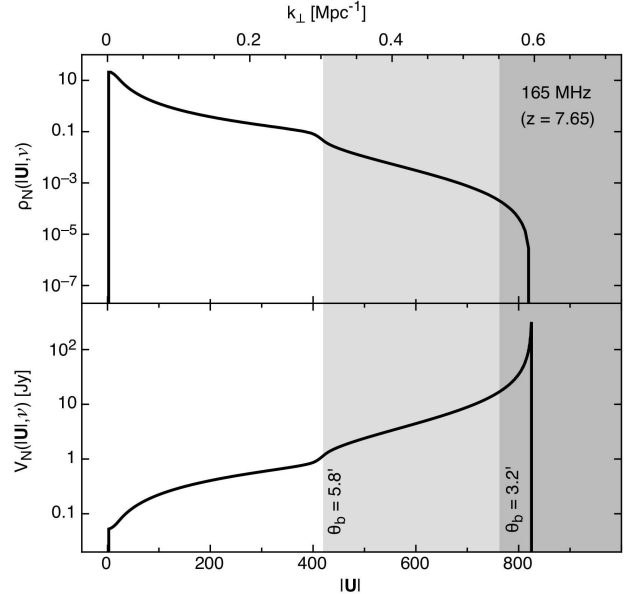} 
\caption{\textit{Top:} Baseline density profile for the MWA configuration described in \S\,4.1 at $\nu \approx 165$\,MHz. \textit{Bottom:} Rotationally invariant system noise per visibility showing truncated visibilities (\textit{shaded regions}) due to variation of synthesized beamwidth ($\theta_{\rm b} = \theta_{\rm FWHM} = 0.705U_{\rm max}^{-1}$ for a circular aperture).}
\label{fig3}
\end{figure}

The MWA will consist of 512 tiles, each with 16 cross-dipoles with 1.07\,m spacing and an effective area $A_{\rm eff}\approx16(\lambda^{2}/4)$ for $\lambda\lsim$ 2.1\,m. The system temperature at $\nu<200$\,MHz is sky dominated and has a value $T_{\rm sys}\sim250\left[(1+z)/7\right]^{2.6}$\,K. Following \cite{bowman2006}, we assume a smooth antenna density profile $\rho_{\rm{ant}}\propto r^{-2}$ within a 750 m radius with a core density of one tile per 18\,m$^{2}$. Due to the large system temperature, weak signal and poor sampling of the largest baselines, naturally weighted visibility data produce, at best, low signal-to-noise images and spectra. In addition, finite $uv$-coverage, determined by the maximum baseline, truncates the simulated visibility data and hence we are unable to use a Gaussian tapering function to suppress small scale sidelobes. In order to maintain the highest sensitivity for spectral fitting (see \S\,4.3), we instead truncate the naturally weighted visibility data using a filled circular aperture of radius $a$ with a corresponding beamwidth\footnote{Note that the `beamwidth' referred to by many authors is $\Delta\theta = 1.22\lambda/D \sim \lambda/D$, which is the highest angular resolution meeting the Rayleigh criterion. This corresponds to the central peak to first null separation in the resulting beam pattern of a circular aperture. Other conventions include: full beam width between first nulls $\theta_{\rm{FWFN}} = 2.44\lambda/D$, full beam width at half power $\theta_{\rm{FWHP}} = 1.02\lambda/D$ and full beam width at half maximum $\theta_{\rm{FWHM}} = 0.705\lambda/D$. We have adopted the last of these conventions in this paper.} (full width at half maximum) $\theta_{\rm b} = 0.705U_{\rm max}^{-1}$, where $U_{\rm max} = a/\lambda$. Figure~\ref{fig3} shows the baseline density profile for the MWA configuration described in \S\,4.1 at $\nu \approx 165$\,MHz, as well as the rotationally invariant system noise per visibility showing truncated visibilities due to variation of synthesized beamwidth. The third and fourth columns of Figure~\ref{fig1} show the resulting synthetic 21\,cm brightness temperature maps of the quasar generated HII regions shown in the second column for beamwidths $\theta_{\rm b}\approx 3.2\,^{\prime}$ and $5.8\,^{\prime}$. For the purposes of imaging, uniform weighting of the visibilities would remove the bias from the greater sampling of short baselines (i.e. low spatial frequencies) and therefore lessen the masking of small scale features in the image.

We simulate the thermal noise in a 3-dimensional visibility-frequency cube using equations (\ref{radiometer}) and (\ref{tU}). This procedure allows us to account for a particular array configuration and observation strategy. We then perform a 2-dimensional inverse Fourier transform in the $uv$-plane for each binned frequency in the bandwidth, which gives a realization of the system noise in the image cube (i.e. sky coordinates). Using the linearity of both the Fourier transform and equation (\ref{vis}), the observed specific intensity can then be found either by first adding visibility terms and then transforming to image space, or by adding the realization of each visibility component in image space. Finally, we apply the Rayleigh-Jeans approximation to express the observed flux in terms of brightness temperature $T$ using $\partial B_{\nu}/\partial T = 2k_{\rm B}/\lambda^{2}$, where $B_{\nu}$ is the specific intensity.

The resulting rms noise in an image constructed in the manner described, for a beamwidth $\theta_{\rm b}$ and frequency bin $\Delta\nu$ has the form
\begin{equation}
\Delta T_{\rm b} \approx T(\theta_{\rm b})\left(\frac{1+z}{8}\right)^{2.6}\left(\frac{\Delta\nu}{100\,\mathrm{kHz}}\frac{t_{\mathrm{int}}}{100\,\mathrm{hr}}\right)^{-1/2}.
\end{equation}
We find that the prefactor takes values of $T(\theta_{\rm b}) \approx 11$\,mK for $\theta_{\rm b} = 3.2\,^{\prime}$ (corresponding to $\Delta\theta = 5.5\,^{\prime}$ central peak to first null) and $T(\theta_{\rm b}) \approx 65$\,mK for $\theta_{\rm b} = 2.9\,^{\prime}$ ($\Delta\theta = 5.1\,^{\prime}$). For comparison, the radiometer equation (e.g. \citeauthor{wlb2005} \citeyear{wlb2005}) estimates the noise within a synthesized beam of width $\theta_{\rm b}$ to have corresponding values of $T(\theta_{\rm b}) \approx 20$ and 22\,mK for $\theta_{\rm b} = 3.2\,^{\prime}$ and 2.9$\,^{\prime}$ respectively. These differences are reconciled by noting that the radiometer equation assumes a \textit{uniform} antenna density $\rho_{\rm{ant}}$, while using equations (\ref{radiometer}) and (\ref{tU}) to calculate the rms noise fluctuation in visibility space accounts for an arbitrary antenna density. For the $\rho_{\rm{ant}} \propto r^{-2}$ antenna density profile of the MWA, the reduced antenna density at large $r$ leads to a reduced baseline density at large $U$ \citep{bowman2006}, and hence increased noise levels at high angular resolution. Figure~\ref{fig3} demonstrates the increased noise levels for the largest values of $U \sim 800$ for the MWA operating at $\nu = 165$\,MHz. In summary, in the case of the MWA, using the radiometer equation (and hence assuming a uniform antenna density) results in an underestimate of noise for large baselines and an overestimate for small baselines.

\begin{figure*}
\includegraphics[width= 17.5cm]{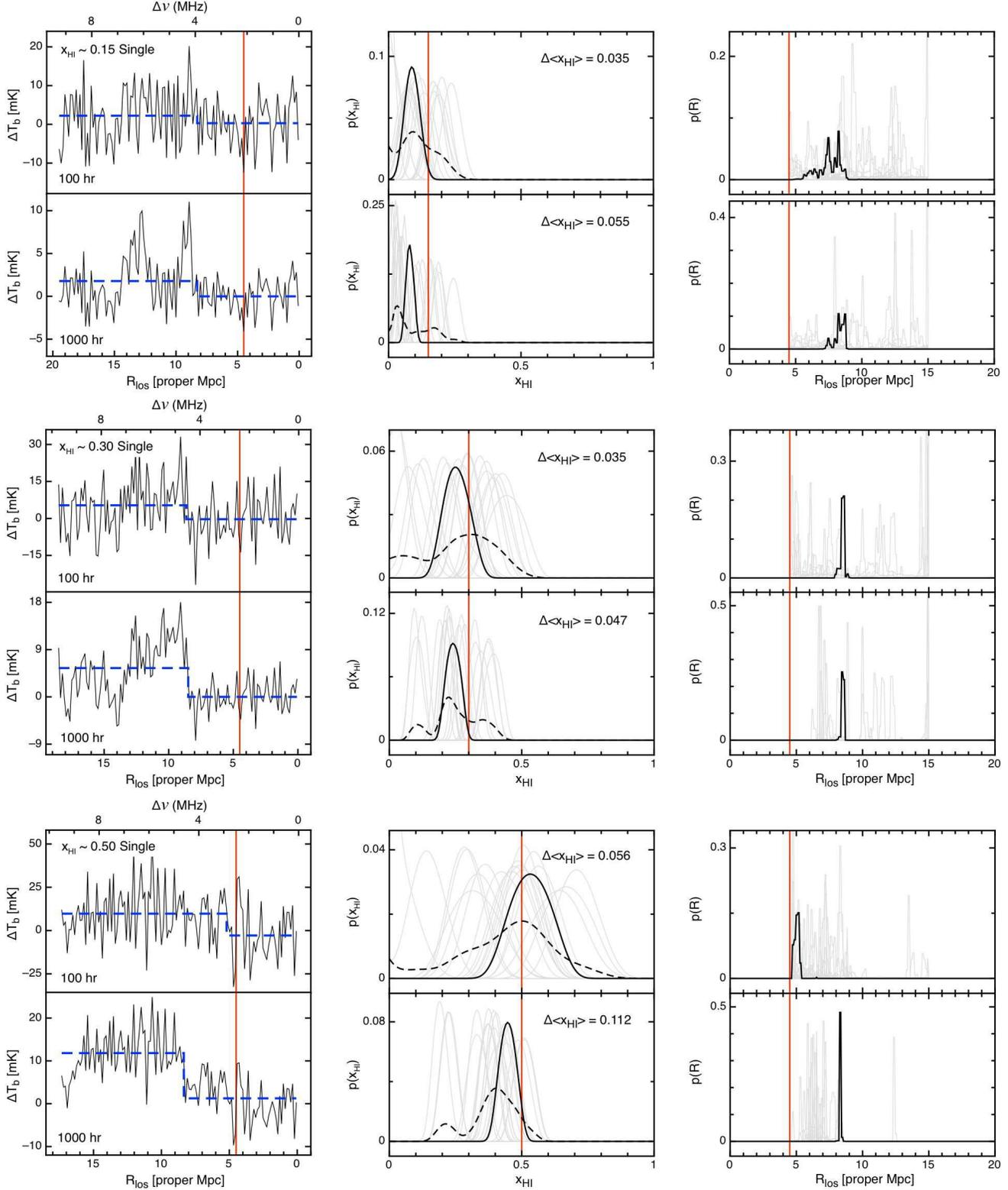} 
\caption{ \textit{Left:} Sample mock spectra of quasar HII regions using the MWA showing the recovered step function (\textit{dashed}). \textit{Center and Right:} Marginalized probability functions of recovered neutral fraction $x_{\rm HI}$ and radius $R$ for an ensemble of simulations (\textit{grey}). One example (\textit{bold}) corresponds to the mock spectrum in the left-hand panel. Likelihood functions $\mathcal{L}(x_{\rm HI}|\langle x_{\rm HI}\rangle)$ for an observed neutral fraction $x_{\rm HI}$ are shown by the dashed curves. The vertical lines show the global neutral fraction of the surrounding IGM and the modeled region radius $R_{\rm q}$. Six cases are shown corresponding to $t_{\rm int} = 100$\,hr and 1000\,hr when $\langle x_{\rm HI}\rangle \approx$ 0.15, 0.3 and 0.5 (\textit{top} to \textit{bottom}). The same IGM realization and system noise (scaled by $1/\sqrt{t_{\rm int}}$) have been used at each redshift to facilitate comparison between integration times.}
\label{fig4}
\end{figure*} 

\subsection{Continuum foreground subtraction}
\label{foregrounds}
Foreground contamination and its removal can have significant consequences for the detectability of the weak, redshifted 21\,cm cosmic signal. Both spatial and frequency characteristics of foreground emission will affect our ability to subtract these contaminants sufficiently in degree and accuracy. For the remainder of this paper we concern ourselves with the detection of HII regions surrounding known high redshift quasars in the spectral regime. We defer discussion of the detectability of such HII regions in the full image regime using a filter matching technique (e.g. \citeauthor{datta2007b} \citeyear{datta2007b}) to future work.

Foreground contamination will be brighter than the cosmological 21\,cm signal by at least four orders of magnitude. We assume that the total foreground signal, which is dominated by galactic synchrotron radiation \citep{shaver1999}, has had extragalactic point sources removed, leaving a spectrally smooth foreground signal that will be removable through continuum subtraction \citep{gnedin2004,wang2006}.

The process of continuum foreground removal will remove any cosmological signal for \textit{line of sight} modes with $k_{\parallel} \lsim$ $2\pi/y$, where $y$ is the depth of the survey \citep{mcquinn2006}. Since HII regions may have sizes corresponding to a significant fraction of the frequency band pass, the insensitivity to large scale modes may hamper the detectability of HII regions. The MWA will operate with a processed bandwidth of 32\,MHz. For comparison, the frequency domain of our synthetic spectra is $\Delta\nu \lsim$ 20\,MHz. Thus, foreground modes along the line of sight with wavelengths longer than the 32\,MHz band pass are not present in our mock spectra, with the exception of an overall constant corresponding to the DC mode of the spectrum.

In this paper we make the assumption that foregrounds could be removed over the 20\,MHz of our simulated spectrum, given access to the 32\,MHz of spectrum that will be available for the MWA.

\subsection{Estimate of signal-to-noise ratio in detection of neutral IGM surrounding HII regions}
\label{SN}
We next present mock spectra of HII regions surrounding known high redshift quasars at times when the global neutral fraction is $\langle x_ {\rm HI}\rangle\approx0.15,$ 0.3 and 0.5, and discuss the prospects for their detection.

Line of sight spectra are simulated using a synthesized beam of width $\theta_{\rm b} = 3.2\,^{\prime}$. Figure~\ref{fig4} shows sample mock spectra for 100\,hr and 1000\,hr integrations centered on 186, 175 and 164\,MHz (corresponding to $z=6.7,$ 7.1 and 7.65 respectively). Instrumental thermal noise in these spectra is generated following the prescription described in \S\,\ref{sensitivity}. The presence of an HII region could be inferred from this data via the detection of a brightness temperature contrast between the ionized IGM within the HII region and the neutral hydrogen outside. We do not know \textit{a priori} the radius of the HII region. However from observations of the proximity zone surrounding a high redshift quasar we know that if an HII region is present, its size must be larger than the redshift corresponding to the observed transmission blueward of the Ly$\alpha$ line in near-IR spectra.

To test whether the mock spectra are of sufficient signal-to-noise to detect an HII region we fit a step function 
\begin{equation}
T_{\rm model}(\nu)=T_{\rm IGM}-\Delta T \times\Theta(R_{\rm region}-R)
\end{equation}
which has three free parameters. These parameters correspond to the brightness temperature outside the region $T_{\rm IGM}$, a radius $R_{\rm region}$ (which we assume to have a value $R>R_{\rm q}$) and a temperature brightness contrast across the HII region boundary $\Delta T$. The parameter $T_{\rm IGM}$ includes the contribution of the DC component of foregrounds which is therefore removed during the fitting procedure. We note that our fit is made to the spectrum in only one pixel of the synthetic map, which includes contributions of foregrounds from all regions within the synthesized beam. Using the calculated thermal noise $\sigma_T$ in each of $N_{\rm ch}$ channels for the MWA with an assumed integration time $t_{\rm int}$, we then calculate the value of $\chi^2$, where 
\begin{equation}
\chi^2 = \sum_{i=1}^{N_{\rm ch}} \left[\frac{T_i-T_{\rm model}(\nu_i)}{\sigma_T}\right]^2
\end{equation}
for each set of parameters, and hence the likelihood 
\begin{equation}
\mathcal{L}(\Delta T,R_{\rm region},T_{\rm IGM})=\exp{\left(-\chi^{2}/2\right)}.
\end{equation}
From Bayes' theorem the \textit{a posteriori} probability for the parameters given the observed spectrum follows from
\begin{eqnarray}
\nonumber
\frac{d^3P}{d\Delta T dR_{\rm region} dT_{\rm IGM}}&&\\
 &&\hspace{-30mm}\propto \mathcal{L}(\Delta T,R_{\rm region},T_{\rm IGM}) \frac{dP_{\rm prior}}{d\Delta T}\frac{dP_{\rm prior}}{dR_{\rm region}}\frac{dP_{\rm prior}}{dT_{\rm IGM}}, 
\end{eqnarray}
where we assume prior probability distributions
\begin{eqnarray}
\nonumber
\qquad\frac{dP_{\rm prior}}{d\Delta T} &\propto& 1,\\
\nonumber
\frac{dP_{\rm prior}}{dR_{\rm region}}&\propto& \Theta(R_{\rm region}-R_{\rm min})\Theta(R_{\rm max}-R_{\rm region}),    \\
\frac{dP_{\rm prior}}{dT_{\rm IGM}} &\propto& 1.
\end{eqnarray}
Here $\Theta$ is the Heaviside step function, $R_{\rm min}=R_{\rm q}$ and $R_{\rm max}=15$ proper Mpc. The value of $R_{\rm max}$ represents an upper limit on the size of a proximity zone within which the quasar ionizing flux is dominant. If the HII region extends beyond $R_{\rm max}$ along a particular line of sight, we effectively count this ionization as part of the general two phase IGM. $R_{\rm min}$ is set by the observed radius of Ly$\alpha$ transmission (indicating a highly ionized IGM). Provided the quasar is beamed with an opening angle $>\alpha$, where $\alpha = \arctan(R_{\rm b}/R_{\rm q})$, this assumption will be valid. This is likely since at $z\approx6.5$ with a synthesized beamwidth of $\sim4^{\prime}$, $\alpha\approx13^\circ$. We then marginalize the probability distribution to obtain constraints on the parameters of interest ($\Delta T$ and $R_{\rm region}$). Finally, the neutral fraction is linearly related to the retrieved temperature contrast through $x_{\rm HI}=\Delta T/\{22\left[(1+z)/7.5\right]^{1/2}\}$\,mK.

The uncertainty in an observation should arise from a combination of IGM fluctuations and instrumental noise. This instrumental noise is thermal and therefore Gaussian, whereas IGM fluctuations are highly non-Gaussian (e.g. \citeauthor{lidz2007a} 2007a; \citeauthor{wm2007} \citeyear{wm2007}). A non-Gaussian noise component renders the $\chi^2$ statistic suboptimal. However, in the case of the MWA, instrumental noise will dominate in narrow frequency channels which justifies the use of $\chi^{2}$ in estimating the likelihood.

\subsubsection{Recovered parameters from mock spectra}
In the central and right-hand panels of Figure~\ref{fig4} we show marginalized \textit{a posteriori} probability density functions for the recovered neutral fraction $x_{\rm HI}$ and radius $R_{\rm region}$ for 100\,hr and 1000\,hr integrations. These parameter estimates correspond to the mock spectra in the left-hand panels of Figure~\ref{fig4}. The vertical lines show the global neutral fraction of the surrounding IGM and the modeled region radius $R_{\rm q}$.

In addition to the telescope noise, the spectra plotted in Figure \ref{fig4} show that the recovered radius will be sensitive to the large fluctuations in brightness  temperature due to the inhomogeneous ionization state of the IGM (\citeauthor{lidz2007b} 2007b). In contrast, the estimates of neutral fraction along a particular line of sight average over many fluctuations for a range of recovered radii. As a result the neutral fraction along a particular line of sight should be measured in an unbiased way. On the other hand, different lines of sight will show a varying systematic offset from the mean due to the inhomogeneous reionization of the IGM.

To investigate the extent of systematic offset due to patchy reionization we have analyzed mock spectra from an ensemble of HII regions and lines of sight. Probability distributions for the best fit parameters $R$ and $x_{\rm HI}$ are estimated from this ensemble and are shown as grey lines in Figure~\ref{fig4}. These distributions quantify the level of systematic uncertainty that is present. In cases where the telescope noise is small, the systematic offsets may be larger than the statistical uncertainty in the neutral fraction. This is demonstrated by the 1000\,hr examples in Figure~\ref{fig4}. The inhomogeneous nature of the IGM, combined with system noise conspire to make the determination of $R$ very uncertain despite high statistical precision.

\begin{figure}
\includegraphics[width=8.7cm]{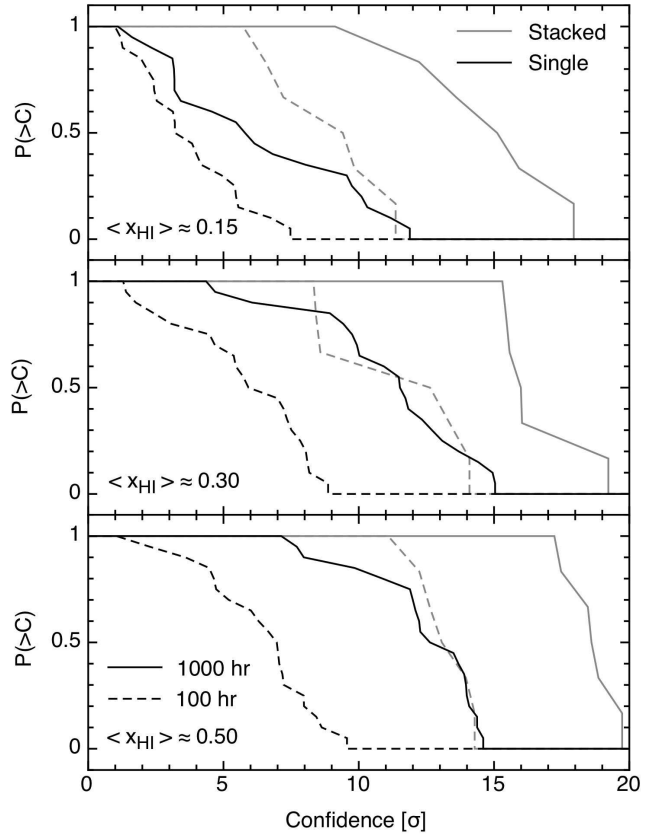}  
\caption{ Cumulative probabilities of detection for 100\,hr (\textit{dashed}) and 1000\,hr (\textit{solid}) integrations using single (\textit{black}) and stacked spectra (\textit{grey}) when the globally averaged neutral fraction is $\langle x_{\rm HI}\rangle \approx$ 0.15, 0.3 and 0.5 (\textit{top} to \textit{bottom}).}
\label{fig5}
\end{figure}

We define the confidence for detection of an HII region $C$ by the ratio of the mean recovered neutral fraction to the standard deviation of the marginalized \textit{a posteriori} probability density functions for the recovered neutral fraction. Cumulative probabilities of detection are calculated using an ensemble of mock line of sight spectra. Figure~\ref{fig5} shows the cumulative probabilities of detection using single line of sight spectra for 100\,hr and 1000\,hr integrations (dark lines). These results indicate that it will be possible to regularly detect bubbles at high significance [$P(C>5)>0.5$] from spectra of individual HII regions using integrations of 1000\,hr, even at redshifts close to overlap when the neutral fraction is as low as $\approx15\%$. However at neutral fractions of $>30\%$, measurement will be possible in only 100\,hr. This is despite the fact that at these late times the typical HII region becomes comparable in size to the HII regions generated by the quasar (at $z\lsim$ 7 in this model). As stressed in \S\,\ref{intro}, even when the impact of the quasar is no longer dominant, the method of detecting the 21\,cm signal \textit{a posteriori} along the line of sight to a known quasar will still succeed, because neutral gas is known \textit{a priori} to be absent in a large volume surrounding the quasar.

\subsection{Extracting neutral fraction from observations}
The discussion in \S\,\ref{SN} describes estimation of the certainty with which an HII region can be detected. Once the existence of neutral gas is established, the interesting question becomes whether the HII region can be used to measure the global neutral fraction. Below we outline how this might be achieved.

The modeling described could be used to generate a likelihood function for an observed neutral fraction $x_{\rm HI}$ given a trial global mean $\langle x_{\rm HI}\rangle$,
\begin{equation}
\mathcal{L}(x_{\rm HI}|\langle x_{\rm HI}\rangle)=\frac{1}{N}\sum_{i=1}^{N}\frac{dP_{i}}{dx_{\rm HI}}\bigg|_{\langle x_{\rm HI}\rangle},
\end{equation}
where $N$ is some large number of model spectra and the $dP_{i}/dx_{\rm HI}$ are the corresponding probability distributions for $x_{\rm HI}$ (e.g. those shown as grey lines in Figure~\ref{fig4}). The resulting likelihood functions for the example neutral fractions are shown in Figure~\ref{fig4} by the dashed curves. These likelihood functions give an idea of the uncertainty on a measurement of $\langle x_{\rm HI} \rangle$ based on an observation of a single HII region. In the case of a 1000\,hr integration, we find the variance of these likelihood functions to be $\sigma_{x_{\rm HI}} \approx 0.07$, 0.09 and 0.1 for $\langle x_{\rm HI}\rangle = 0.15$, 0.3 and 0.5 respectively. The corresponding relative errors $\sigma_{x_{\rm HI}}/\langle x_{\rm HI} \rangle$ are 0.5, 0.3 and 0.2. These values represent estimates of the achievable uncertainty.

Given an observed spectrum with probability $dP_{\rm obs}/dx_{\rm HI}$ for $x_{\rm HI}$ along that line of sight, it would then be possible to find the \textit{a posteriori} probability for the globally averaged neutral fraction $\langle x_{\rm HI}\rangle$,
\begin{equation}
\frac{dP}{d\langle x_{\rm HI}\rangle}\bigg|_{\rm obs} = \int\frac{dP_{\rm obs}}{dx_{\rm HI}}\mathcal{L}(x_{\rm HI}|\langle x_{\rm HI}\rangle)\frac{dP_{\rm prior}}{d\langle x_{\rm HI}\rangle},
\end{equation}
where $dP_{\rm prior}/d\langle x_{\rm HI}\rangle$ is the prior probability for the neutral fraction along that line of sight.

\subsection{Recovered parameters from stacked spectra of a number of HII regions}
In the previous section we presented results for mock observations of single HII regions like those expected to be present around luminous high redshift quasars. The observed number density of quasars with luminosities greater than $L$ scales as $N(>L)\propto L^{-2}$ \citep{richards2006}. Thus while current surveys only detect $\sim1$ luminous quasar per MWA observing field, in the future, deeper surveys will uncover less luminous high redshift quasars at greater number densities. This increase in density towards lower luminosity will make it possible to stack multiple spectra, thereby reducing fluctuations in the ionization state of the IGM along particular lines of sight which cannot be removed through a long integration time \citep{wlb2005}. In this section we investigate whether an increased sensitivity for the detection of contrast between neutral and ionized regions could be obtained by stacking spectra of smaller HII regions around a number of less luminous high redshift quasars. We consider a value of $R_{\rm q}\approx21$ comoving Mpc, which may be more appropriate for a fainter population of quasars than the 34 comoving Mpc we assume for the single HII regions studied in the previous subsection.  

Mock spectra are simulated as before for 10 HII regions surrounding host halos of minimum mass $10^{12}$\,$M_{\odot}$, and then stacked to obtain a single averaged spectrum. The stacked spectra are more representative of the average IGM surrounding quasars and have decreased levels of instrumental noise. We find that averaging over the lines of sight to many quasars greatly reduces the systematic component of uncertainty in the neutral fraction introduced by variation among individual lines of sight. We fit step functions to the mock stacks of observed spectra with three free parameters and estimate the probability distributions of best fit parameters from an ensemble of simulations as before. Figure~\ref{fig5} shows the cumulative probabilities for detection using stacked spectra from 100\,hr and 1000\,hr integrations (grey lines). These results indicate that it will be possible to detect bubbles at high significance [$P(C>5)>0.5$] from stacked spectra, even at redshifts close to overlap and with modest integration times of $\sim100$\,hr.

\section{Discussion}
\label{conclusion}
We have analyzed the prospects for detection of HII regions surrounding luminous high redshift quasars in a percolating IGM prior to the end of the epoch of reionization. We find that fluctuations in the HI distribution of the surrounding IGM will limit the accuracy (though not the precision) with which the global neutral fraction may be recovered owing to the finite length of the observed spectra. Thus the fluctuating nature of the IGM will lower the signal-to-noise ratio achievable on the detection of neutral gas in the IGM surrounding high redshift quasars. This degradation of the signal-to-noise becomes more serious as reionization proceeds since the impact of the quasar on the ionization of the surrounding IGM becomes progressively less significant. At these late times in the reionization epoch, the fluctuations in the ionization structure of the IGM that accompany the percolation of galaxy HII regions will also significantly affect the morphology of HII regions surrounding the high redshift quasars, leading to the formation of complicated, non-spherical boundaries for isotropic sources.

Nevertheless, even at late times when the global hydrogen neutral fraction is as low as $\langle x_{\rm HI}\rangle\approx15\%$ the large volume of ionized gas which is known \textit{a priori} to surround the quasar will imprint a detectable signature on the observed spectrum that is distinct from a region of typical IGM. We find that the presence of neutral gas surrounding a single HII region centered on a luminous high redshift quasar will be detectable in as little as 100 hours ($\langle x_{\rm HI}\rangle\gsim\;30\%$) using the MWA in over 50\% of cases at a significance greater than $5\sigma$. A highly significant detection of neutral gas will be possible in 100\,hr for a stack of 10 smaller 3 proper Mpc HII regions. In both cases an accurate determination of the value of neutral fraction will be hampered by lack of knowledge regarding the detailed shape of the HII region. We estimate that fluctuations in the overlapping IGM between different lines of sight will limit the accuracy with which the global neutral fraction can be inferred from a single HII region to $\sim50\%$, 30\% and 20\% for $\langle x_{\rm HI}\rangle \approx$ 0.15, 0.3 and 0.5 respectively.

The results of this study imply that the detection of neutral IGM surrounding high redshift quasars could be achieved utilizing either a single luminous quasar, or the stacked observations of a number ($\sim$\,10) of less luminous quasars. Obtaining a sufficient number of high redshift quasars within a single MWA field to allow for the possibility of stacking spectra will require an optical/near-IR survey for high redshift quasars in the southern sky with flux limits around half a magnitude fainter than existing northern sky surveys \citep{wlb2005}. Several suitable surveys for high redshift in the southern sky are currently planned (e.g. SkyMapper\footnote{http://www.mso.anu.edu.au/skymapper/index.php} \& VISTA\footnote{http://www.vista.ac.uk/}). The direct observation of neutral gas surrounding the HII regions of high redshift quasars discovered in these surveys would be complimentary to the statistical detection of an ensemble of HII regions via measurement of the power spectrum of 21\,cm fluctuations (e.g. \citeauthor{tozzi2000} \citeyear{tozzi2000}; \citeauthor{furl2004b} \citeyear{furl2004b}; \citeauthor{loeb2004} \citeyear{loeb2004}; \citeauthor{wm2007} \citeyear{wm2007}; \citeauthor{iliev2006} \citeyear{iliev2006}), and would represent a direct detection of the end of the reionization era. 

{\bf Acknowledgments} We thank Bob Sault for invaluable conversations regarding interferometric noise simulations and the anonymous referee for providing useful and detailed suggestions for improving the original manuscript. We also thank Lister Staveley-Smith for helpful comments on a draft of the manuscript. PMG is grateful for many insightful discussions with Lila Warszawski and acknowledges the support of an Australian Postgraduate Award. The research was supported by the Australian Research Council (JSBW).

\newcommand{\noopsort}[1]{}

\bibliography{bib}

\label{lastpage}
\end{document}